\documentclass[10pt,twocolumn,letterpaper]{article}

\usepackage[pagenumbers]{cvpr} 

\usepackage{graphicx}
\usepackage{amsmath}
\usepackage{amssymb}
\usepackage{booktabs}
\usepackage[dvipsnames]{xcolor}
\usepackage{multirow}
\usepackage{hhline}
\usepackage{mathtools}
\usepackage{siunitx}
\usepackage{array}
\usepackage{tabularx}
\usepackage{lipsum}

\newcolumntype{C}[1]{>{\centering\arraybackslash}m{#1}}

\DeclarePairedDelimiter{\norm}{\lVert}{\rVert}

\usepackage[pagebackref,breaklinks,colorlinks]{hyperref}

\usepackage[capitalize]{cleveref}
\crefname{section}{Sec.}{Secs.}
\Crefname{section}{Section}{Sections}
\Crefname{table}{Table}{Tables}
\crefname{table}{Tab.}{Tabs.}

\def\cvprPaperID{3661} 
\def\confName{CVPR}
\def\confYear{2022}

\begin{document}

\title{LC-FDNet: Learned Lossless Image Compression with Frequency Decomposition Network}

\author{Hochang~Rhee$^1$, Yeong~Il~Jang$^1$, Seyun~Kim$^2$, Nam~Ik~Cho$^1$\\
$^1$Seoul National University, Seoul, Korea\\
$^1$Dept. of Electrical and Computer Engineering, INMC\\
$^2$Gauss Labs\\
{\tt\small {hochang,jyicu}@ispl.snu.ac.kr, light4u@gmail.com, nicho@snu.ac.kr}}

\maketitle

\begin{abstract}
Recent learning-based lossless image compression methods encode an image in the unit of subimages and achieve comparable performances to conventional non-learning algorithms. However, these methods do not consider the performance drop in the high-frequency region, giving equal consideration to the low and high-frequency areas. In this paper, we propose a new lossless image compression method that proceeds the encoding in a coarse-to-fine manner to separate and process low and high-frequency regions differently. We initially compress the low-frequency components and then use them as additional input for encoding the remaining high-frequency region. The low-frequency components act as a strong prior in this case, which leads to improved estimation in the high-frequency area. In addition, we design the frequency decomposition process to be adaptive to color channel, spatial location, and image characteristics. As a result, our method derives an image-specific optimal ratio of low/high-frequency components. Experiments show that the proposed method achieves state-of-the-art performance for benchmark high-resolution datasets.
\end{abstract}

\section{Introduction}
\label{sec:introduction}

\begin{figure}[t]
    \centering
    \includegraphics[width=1.0\linewidth]{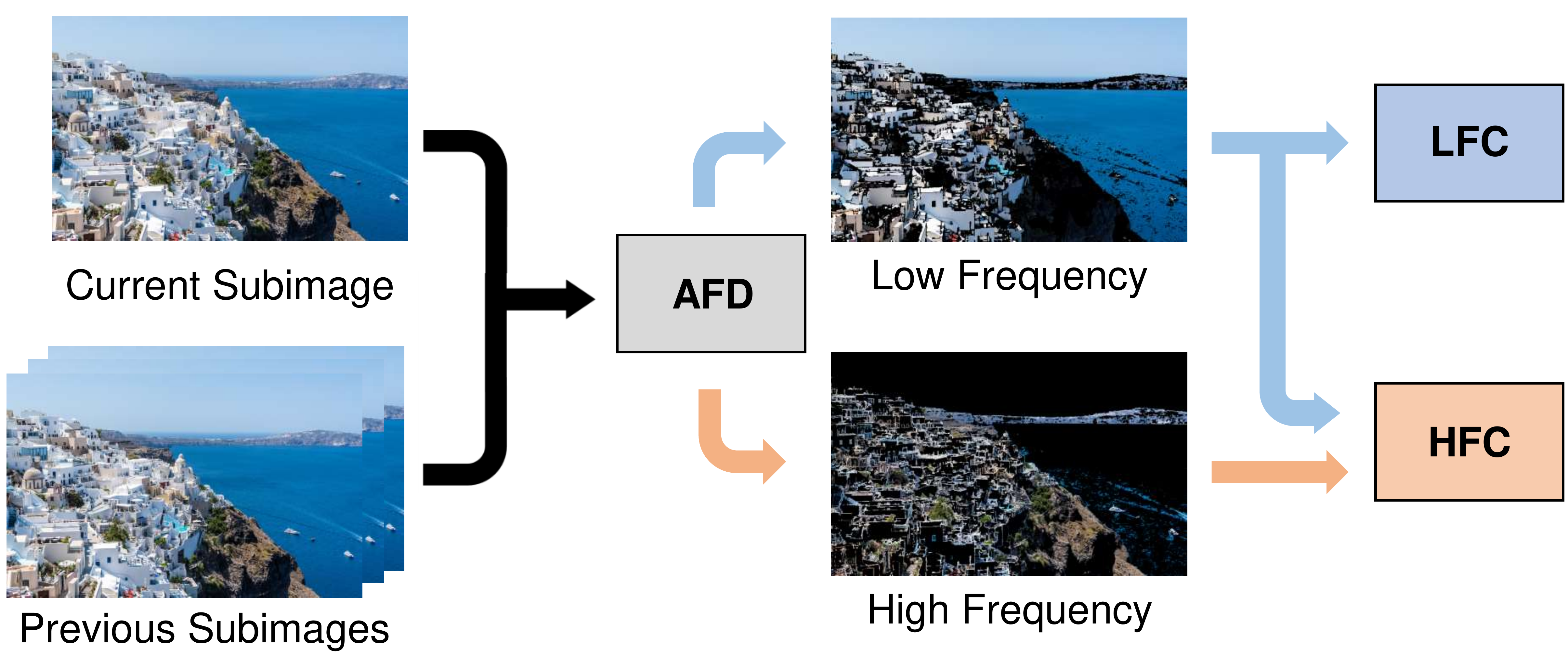}
    \caption{Our LC-FDNet consists of Adaptive Frequency Decomposition (AFD), Low-Frequency Compressor (LFC), and High-Frequency Compressor (HFC). The current subimage is split into low/high-frequency regions through the AFD. The LFC first compresses the low-frequency region, and then the HFC compresses the high-frequency area using the low as strong prior.}
    \label{fig:teaser}    
\end{figure}

As the need for high-quality images is increasing, the importance of image compression is growing accordingly. Driven by the development of deep neural networks (DNNs), there has been remarkable progress in computer vision and image processing, including lossy~\cite{toderici_rnn, balle_end2end, thesis_lossy, toderici_full_rnn, rippel_realtime, Hyperprior, hierarchicalprior, MentzerLossy, contextadaptive, conditional_ae, gaussian_mixture, content_weighted, wavelet-like, EnergyCompaction, mentzer_highfidelity, nearlossless, CWIC, lossyprior, lossy_transferability, lossy_attention, lossy_checkerboard, lossy_slimmable, lossy_asymmetric, lossy_homography, lossy_variablerate, lossy_attentional} and lossless image compression~\cite{PixelCNN, PixelCNN++, MSPixelCNN, PixelRNN, L3C, Lossy2Lossless, nearlossless, Rhee_Access, idf, CBPNN}. Although lossy compression is generally preferred, lossless compression is also necessary for many applications. Lossless compression is especially required for medical images, scientific images, technical drawings, and artistic photos. While methods such as JPEG2000 (lossless mode)~\cite{JPEG2000} employ transform coding with discrete wavelet transform (DWT), most of the standard/non-standard lossless compression methods~\cite{BPG, PNG, JPEG-LS, JPEG-XL} use predictive coding. The standard predictive coding scheme uses a closed-loop prediction where the current pixel is estimated and compressed using the previously encoded samples.

In this sense, early learning-based lossless compression algorithms~\cite{PixelRNN, PixelCNN, PixelCNN++, MSPixelCNN, L3C, Lossy2Lossless} design DNNs as autoregressive models. They rely on the strong power of DNNs in estimating the probability distribution of a pixel conditioned on the previous samples. For example, PixelRNN~\cite{PixelRNN}, PixelCNN~\cite{PixelCNN}, and PixelCNN++~\cite{PixelCNN++} compress each pixel sequentially, where the probability distribution is predicted conditioned on all previous pixels. However, these methods require neural network computations for the number of whole pixels, leading to an impractical inference time.

To achieve practicality, recent works~\cite{MSPixelCNN, L3C, Lossy2Lossless} process the encoding in the unit of an entire image or subimages rather than individual pixels. These methods derive the probability distribution of a subimage conditioned on the previously encoded subimages, or the distribution of a whole image conditioned on the lossy compressed image. They show reduced and practical computation time compared to pixel-wise encoding methods. However, these methods consider the low and high-frequency regions equally, giving the same encoding strategies to the regions of different characteristics. In general, it is difficult to obtain optimal performance in high-frequency regions near an edge or texture where the pixel values change rapidly.

We address this challenge and propose Lossless Compression with Frequency Decomposition Network (LC-FDNet) illustrated in Fig.~\ref{fig:teaser}, which consists of Adaptive Frequency Decomposition (AFD), Low-Frequency Compressor (LFC), and High-Frequency Compressor (HFC). We also decompose an image into subimages based on our unique decomposition scheme, and the first subimage is compressed by a conventional lossless compressor. Then, the rest subimages are sequentially compressed by  Fig.~\ref{fig:teaser}. Using the previously encoded and current subimages as the input, the AFD decomposes the image into low and high-frequency regions, and the compressors (LFC and HFC) encode low and high-frequency regions differently. Since the low-frequency region is typically well predicted, we first compress the low-frequency components. On the other hand, high-frequency regions usually exhibit relatively large prediction errors, and hence we encode them separately with additional priors, which are the encoded low-frequency pixels. That is, we feed the low-frequency components as additional input for compressing the high-frequency region.

For the image-specific frequency decomposition, the AFD generates \textit{error variance map} and \textit{error variance thresholds}. Error variance map can be comprehended as the magnitude of the prediction error produced by the network. By thresholding the error variance map with the error variance threshold, we can classify the pixels into low and high-frequency ones. Since the error variance differs depending on the channel, spatial location, and image characteristics, we design the threshold to be adaptive to those factors. This drives the frequency decomposition process to be image-specific, where different threshold values are derived depending on the image property. Experiments show that the proposed method achieves state-of-the-art performance for benchmark high-resolution datasets with reasonable inference time.

In summary, the main contributions are as follows:

\begin{itemize}
\item We propose a lossless image compression framework that compresses in a coarse-to-fine manner, using the low frequency components to boost the performance in high-frequency regions.

\item We design the frequency decomposition process to be adaptive to channel, spatial location, and image characteristics. Hence, the encoding becomes image-specific, improving the compression performance.

\item Our method achieves state-of-the-art performance for benchmark high-resolution datasets with reasonable inference time.
\end{itemize}

\section{Related Works}
\label{sec:related_works}
\noindent\textbf{Pixel-wise Lossless Compression} Learning-based lossless compression methods generally adopt an autoregressive model. Early methods proceeded the encoding in the pixel unit, where each pixel is compressed based on the previously encoded ones. For example, PixelRNN~\cite{PixelRNN} and PixelCNN~\cite{PixelCNN} modeled a pixel as the product of conditional distributions $p(\textbf{x})=\prod_{i}p(x_i\vert x_1,...,x_{i-1})$, where $x_i$ is a single pixel. PixelCNN++~\cite{PixelCNN++} was proposed as an advancement of the above works and achieved performance enhancement along with faster time. They modeled the pixels as a discretized logistic mixture likelihood, used downsampling to capture structure at multiple resolutions, and introduced additional short-cut connections. Despite these factors, PixelCNN++ still maintains the inherent limitation of autoregressive models, {\em i.e.}, network computation is required for each pixel, requiring impractical inference time.

\begin{figure*}[t]
    \centering
    \includegraphics[width=1.0\linewidth]{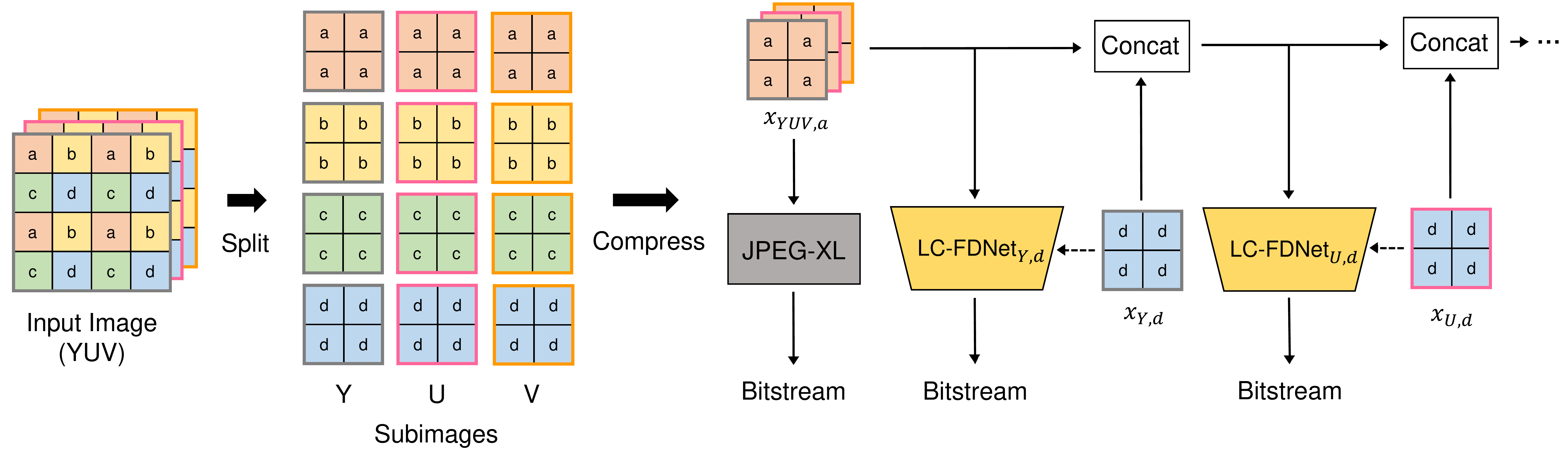}
    \caption{The framework of our compression scheme. Depending on the spatial location, each pixel is grouped as either $a,b,c,d$. The input image is split into subimages, which are sequentially compressed. The subimage $x_{YUV,a}$ is initially encoded using a conventional compression algorithm. The remaining subimages are compressed through deep networks, which receive the previously encoded subimages as input and compress the current subimage. The dotted arrow denotes that the corresponding subimage is currently being compressed. The compressed subimage is then used as an additional input for encoding the next subimage.}
    \label{fig:framework}    
\end{figure*}

\vspace{0.2cm}
\noindent\textbf{Subimage-wise Lossless Compression} For the lossless compression in a reasonable time, recent works perform the encoding in the unit of an entire image or subimages. Each of these methods has its unique strategy for converting an image into subimages. MS-PixelCNN~\cite{MSPixelCNN} first proposed a parallelized PixelCNN using a hierarchical encoding scheme. Specifically, the input image is explicitly divided into four subimages depending on the spatial location, and the distribution of a subimage is conditioned on the previously encoded subimages. However, they used PixelCNNs for modeling the dependency between the subimages, which still required impractical time. L3C~\cite{L3C} proposed a practical compression framework that utilizes a hierarchical probabilistic model. The subimages are implicitly modeled by a neural network and each subimage is conditioned on the subimage of the previous scale. Here, the initial subimage is assumed as a uniform distribution. RC~\cite{Lossy2Lossless} can be seen as a method that divides the image into two parts: lossy compressed image and its residuals. The probability distribution of the residuals is modeled based on lossy compression.

\section{Method}
\label{sec:method}

\subsection{Overview}
\label{sec:overview}
The overall procedure of our method is illustrated in Fig.~\ref{fig:framework}. Given the input image $x \in \mathcal{R}^{H \times W \times 3}$, we first convert the RGB image into a YUV format through a reversible color transform~\cite{RCT}. Then we split the image in a channel-wise and spatial-wise manner. Specifically, we divide the input image into 12 subimages $x_{c,s} \in \mathcal{R}^{\frac{H}{2} \times \frac{W}{2} \times 1}$, where $c$ denotes the channel index ($c\in\{Y,U,V\}$) and $s$ denotes the spatial location index ($s\in\{a,b,c,d\}$). The subimages $x_{YUV,a} = \{x_{Y,a}, x_{U,a}, x_{V,a}\}$ are first compressed using a conventional compression algorithm. Then, the remaining subimages are compressed one by one with our LC-FDNet, where previously encoded subimages are used as input. The order of the subimages to be compressed will be explained in Sec.\ref{sec:split}.

\subsection{Reversible Color Transform}
\label{sec:rct}
In general, RGB images have significant correlations between the color channels. Most standard image/video compression methods adopt YUV transformation to decorrelate the color channels and enhance the compression efficiency. In the case of lossless compression, the YUV transformation must be itself lossless, where the inverse of the YUV back to the RGB should be lossless in integer arithmetic. In this paper, we adopt the reversible color transform proposed in \cite{RCT} since it well approximates the conventional YUV transformation. Note that the $Y$ channel is expressed in 8 bits and $UV$ channels are expressed in 9 bits.

\begin{figure*}[t]
    \centering
    \includegraphics[width=0.85\linewidth]{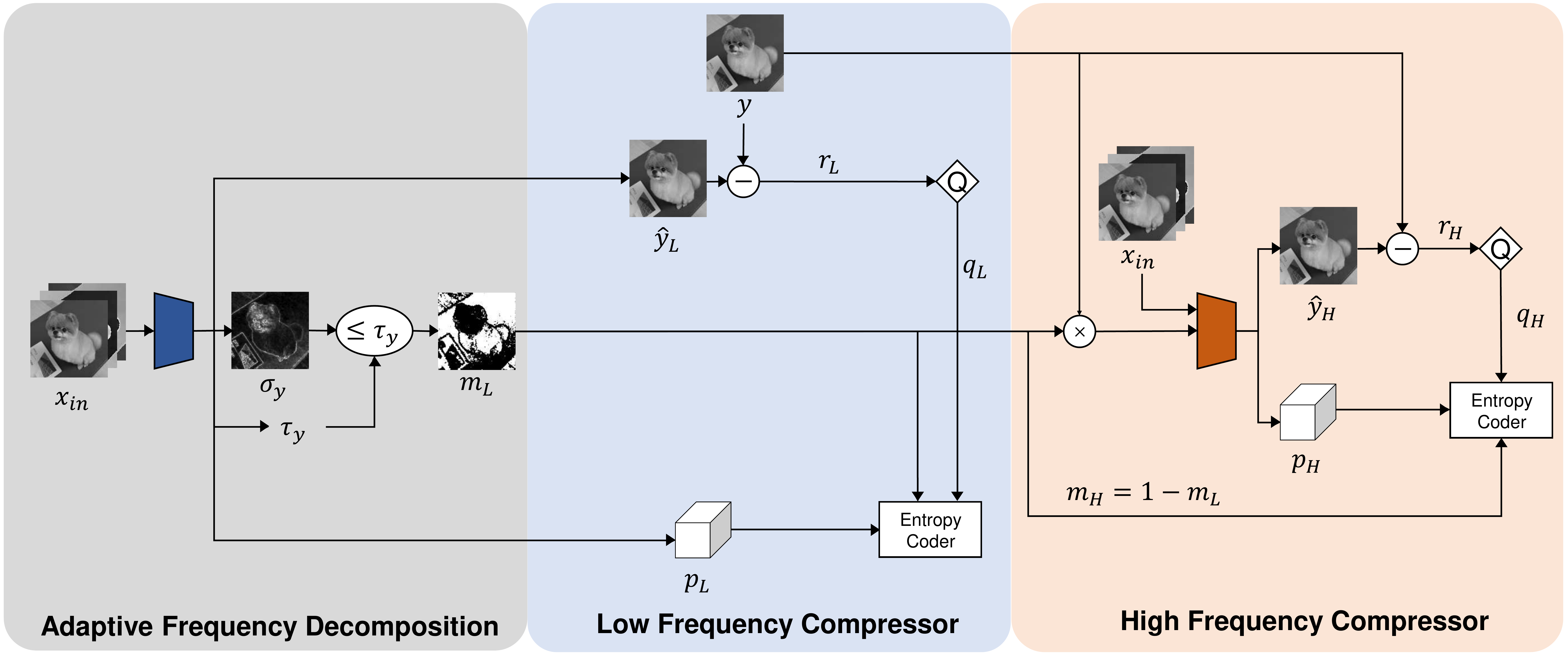}
    \caption{The architecture of LC-FDNet. In this figure, we consider the case of compressing $y=x_{Y,d}$ given $x_{in}=x_{YUV,a}$. AFD part first receives $x_{in}$ and determines each pixel as belonging to either low or high-frequency regions, using error variance map $\sigma_y$ and error variance threshold $\tau_y$. Afterward, LFC encodes the low-frequency region of subimage $y$. HFC then receives the encoded low-frequency region as additional input and compresses the remaining high-frequency region. The decoding process is provided in the Supplementary Material.}
    \label{fig:architecture}    
\end{figure*}

\subsection{Framework}
\label{sec:split}
After the reversible color transformation, the input image is divided into subimages depending on the color channel and the spatial location. Fig.~\ref{fig:framework} shows how we categorize the pixels into four groups ($a,b,c,d$) depending on the spatial location. Pixels in the odd row and odd column are categorized as $a$, odd row and even column as $b$, and so on. 

We compress 12 subimages in total, where the compression of each subimage is conditioned on the previously encoded subimages. To be specific, for the compression of the $N$-th subimage $y \in \mathcal{R}^{\frac{H}{2} \times \frac{W}{2} \times 1}$, we concatenate the already encoded $N-1$ subimages and use it as the input, which we denote as $x_{in} \in \mathcal{R}^{\frac{H}{2} \times \frac{W}{2} \times({N-1})}$. We neglect the notation of the subimage index $N$ for the sake of simplicity. In this scenario, the order of the subimages is critical to computational efficiency. The compression performance is improved as the correlation among the input and the $N$-th subimage increases. For instance, encoding subimage $x_{Y,d}$ is much easier when it is conditioned on $x_{Y,a}$ rather than $x_{V,b}$.

We design the order of the subimages considering the following two factors : 1) color channel and 2) spatial location. In terms of the color channel, we arrange the order as $Y \xrightarrow{} U \xrightarrow{} V$. This is a straightforward choice since the $Y$ channel contains more significant features than $U$ and $V$. Considering the spatial location, we design the network to proceed in the order of $a \xrightarrow{} d \xrightarrow{} b \xrightarrow{} c$. We figure that this is a better design choice compared to MS-PixelCNN~\cite{MSPixelCNN}, where they progress in the order of $a \xrightarrow{} b \xrightarrow{} c \xrightarrow{} d$. Comparing $a \xrightarrow{} d$ and $a \xrightarrow{} b$, we see that $d$ fully utilizes the information of $a$ both horizontally and vertically. In contrast, acquiring $b$ conditioned on $a$ may have more benefit in terms of the horizontal axis, but lacks to fully utilize the vertical components. In conclusion, we proceed the compression of the subimages in the order of $x_{Y,a} \xrightarrow{} x_{U,a} \xrightarrow{}, x_{V,a} \xrightarrow{} x_{Y,d} \xrightarrow{} x_{U,d} \xrightarrow{} ... \xrightarrow{} x_{V,c}$.

For the compression of the initial subimage we adopt a conventional lossless compression algorithm, similar to RC~\cite{Lossy2Lossless}. Prior works~\cite{L3C, Hyperprior} provide the initial prior as uniform distribution or unit Gaussian distribution. Although DNNs show great strength in estimating conditional probability distributions, the strength is limited when weak priors are given. Conventional algorithms instead show competitive performance in this environment, and hence we use them for compressing the initial subimage $x_{YUV,a}$. Specifically, we adopt JPEG-XL~\cite{JPEG-XL} which yields state-of-the-art performance among the conventional compression algorithms.

\subsection{Architecture}
\label{sec:architecture}
In this section, we present the architecture of LC-FDNet illustrated in Fig. \ref{fig:architecture}, which shows the details of Fig.~\ref{fig:teaser}. The goal is to compress the $N$-th subimage $y$ given the input $x_{in}$, which is the concatenation of $N-1$ previous subimages. Note that LC-FDNet is required for each of the subimages, resulting in 9 LC-FDNets in total (since 3 subimages are encoded by JPEG-XL). These networks do not share the parameters since each of them is specific for each subimage. 
 
Throughout the paper, notations $L$ and $H$ denote low and high-frequency, respectively. We first explain the notations in the AFD and LFC parts.

\vspace{0.2cm}
\noindent \textbf{Subimage Prediction} $\hat{y}_{L} \in \mathcal{R}^{\frac{H}{2} \times \frac{W}{2} \times 1}$ is the network prediction of $y$, where better prediction yields more compact compression. Given the prediction, the residual is computed as $r_{L} = \hat{y}_{L} - y$, which is the difference between the ground truth subimage and the prediction. Since the obtained residual is not in the form of integers, we quantize the residual. It is denoted as $q_L$, which is then passed to the entropy coder.

\vspace{0.2cm}
\noindent \textbf{Probability Distribution} $p_L$ is the estimated probability distribution of the quantized residual $q_L$. We directly estimate the probability distribution as the probability mass function (pmf). Hence, the dimension of $p_L$ is $\frac{H}{2} \times \frac{W}{2} \times C$, where $C$ is 511 for the $Y$ channel, and 1021 for the $U,V$ channel. Softmax operation is applied before deriving $p_L$, so that the probabilities sum up to 1.

\vspace{0.2cm}
\noindent \textbf{Error Variance Map} The error variance map $\sigma_{y}\in \mathcal{R}^{\frac{H}{2} \times \frac{W}{2} \times 1}$ represents the estimation of the prediction error magnitude generated by the network. We design the error variance map to follow the magnitude of the prediction error through the following loss:

\begin{equation}\label{eq:loss_errorvariance}
L_{ev} = \norm{\sigma_{y} - |y - \hat{y}_L|}_{1}.
\end{equation}

\noindent Here, each value in the map can be interpreted as the prediction error variance at the corresponding pixel. A large value implies that the network is likely to make a large prediction error at the point, which means that the pixel belongs to a high-frequency region. Similarly, smooth regions \textit{i.e.}, low-frequency regions yield low error variance values.

\vspace{0.2cm}
\noindent \textbf{Error Variance Threshold} With the obtained error variance map, we apply a simple thresholding to categorize each pixel into two classes; pixels in low or high-frequency regions. However, the threshold value should be adjusted depending on the channel, spatial location and image characteristics. For instance, the error variance is typically larger in the $Y$ channel compared to $U$ and $V$. Thus the threshold should be larger in the $Y$. Therefore, instead of a fixed threshold, we let the network derive a specific error variance threshold $\tau_{y} \in \mathcal{R}$ for each subimage. Note that 9 threshold values are derived for a single input image. With $\sigma_y$ and $\tau_{y}$ in hand, we acquire the low-frequency mask as 

\begin{equation}\label{eq:mask}
m_L^i =\begin{dcases}
1 & \text{if $\sigma_y^i \leq \tau_{y}$}
\\
0 & \text{else},
\end{dcases}
\end{equation}

\noindent where $i$ denotes the pixel index. $m_L$ serves as an indicator of which components are considered in the low-frequency region.

\vspace{0.2cm}
The quantized residual $q_L$, corresponding probability distribution $p_L$, and the low-frequency mask $m_L$ are passed to the entropy coder. We compress only the low-frequency components \textit{i.e.}, pixels corresponding to $m_L^i=1$. It can be assumed that pixels belonging to low-frequency regions will have marginal performance enhancement even when additional information is given, specifically when it is compressed in the HFC. Instead, the compression efficiency gain is significant when these components serve as the additional input.

After the compression of low-frequency regions in LFC, HFC encodes the remaining high-frequency regions. Besides $x_{in}$, HFC additionally receives the low-frequency component of the currently encoding subimage $y \odot m_L$. From the input, HFC generates the following two outputs: 1) $\hat{y}_H$ : the prediction of $y$, 2) $p_H$ : the probability distribution of the quantized residual $q_H$. Since low-frequency components serve as a strong prior for the high-frequency components, HFC can make more precise predictions. In addition, the variance of the probability distribution is reduced, leading to compression efficiency. 

The pipeline of HFC is similar to that of LFC. The quantized residual $q_H$, corresponding probability distribution $p_H$, and the high-frequency mask $m_H=1-m_L$ are fed to the entropy coder. Note that HFC can ignore the estimation of low-frequency components and only focus on the high-frequency ones.

\begin{table*}
\centering
\caption{Comparison of our method with other non-learning and learning-based codes on high-resolution benchmark dataset. We measure the performances in bits per pixel (bpp). Best performance is highlighted in bold and the second-best performance is denoted with $^{*}$. The difference in percentage to our method is highlighted in green.}
\resizebox{1.5\columnwidth}{!}{%
\begin{tabular}{clll}
\hline
\hspace{0.5cm} Method \hspace{0.5cm} & \hspace{0.5cm} CLIC.m \hspace{0.5cm} & \hspace{0.5cm} CLIC.p \hspace{0.5cm} & \hspace{0.5cm} DIV2K \hspace{0.5cm} \\ \hline \hhline{====}

PNG \cite{PNG}                      & \hspace{0.5cm} 11.79 \scriptsize\textcolor{OliveGreen}{+69.2\%} \hspace{0.5cm} & \hspace{0.5cm} 11.79 \scriptsize\textcolor{OliveGreen}{+49.2\%} \hspace{0.5cm} & \hspace{0.5cm} 12.69 \scriptsize\textcolor{OliveGreen}{+55.3\%} \hspace{0.5cm}\\
JPEG-LS \cite{JPEG-LS}            & \hspace{0.5cm} \:\:7.59 \scriptsize\textcolor{OliveGreen}{\:\:+8.9\%} \hspace{0.5cm}  & \hspace{0.5cm} \:\:8.46 \scriptsize\textcolor{OliveGreen}{\:\:+7.1\%} \hspace{0.5cm}  & \hspace{0.5cm} \:\:8.97 \scriptsize\textcolor{OliveGreen}{\:\:+9.8\%} \hspace{0.5cm}\\
JPEG2000 \cite{JPEG2000}          & \hspace{0.5cm} \:\:8.13 \scriptsize\textcolor{OliveGreen}{+16.6\%} \hspace{0.5cm}  & \hspace{0.5cm} \:\:8.79 \scriptsize\textcolor{OliveGreen}{+11.3\%} \hspace{0.5cm}  & \hspace{0.5cm} \:\:9.36 \scriptsize\textcolor{OliveGreen}{+14.6\%} \hspace{0.5cm}\\
WebP \cite{WebP}                  & \hspace{0.5cm} \:\:8.19 \scriptsize\textcolor{OliveGreen}{+17.5\%} \hspace{0.5cm}  & \hspace{0.5cm} \:\:8.70 \scriptsize\textcolor{OliveGreen}{+10.1\%} \hspace{0.5cm}  & \hspace{0.5cm} \:\:9.33 \scriptsize\textcolor{OliveGreen}{+14.2\%} \hspace{0.5cm}\\
BPG \cite{BPG}                    & \hspace{0.5cm} \:\:8.52 \scriptsize\textcolor{OliveGreen}{+22.2\%} \hspace{0.5cm}  & \hspace{0.5cm} \:\:9.24 \scriptsize\textcolor{OliveGreen}{+17.0\%} \hspace{0.5cm}  & \hspace{0.5cm} \:\:9.84 \scriptsize\textcolor{OliveGreen}{+20.4\%} \hspace{0.5cm}\\
FLIF \cite{FLIF}                  & \hspace{0.5cm} \:\:7.44 \scriptsize\textcolor{OliveGreen}{\:\:+6.7\%} \hspace{0.5cm}  & \hspace{0.5cm} \:\:8.16 \scriptsize\textcolor{OliveGreen}{\:\:+3.3\%} \hspace{0.5cm}  & \hspace{0.5cm} \:\:8.73 \scriptsize\textcolor{OliveGreen}{\:\:+6.9\%} \hspace{0.5cm}\\ 
JPEG-XL \cite{JPEG-XL}            & \hspace{0.5cm} \:\:7.20$^{*}$ \scriptsize\textcolor{OliveGreen}{+3.3\%} \hspace{0.5cm}  & \hspace{0.5cm} \:\:8.19 \scriptsize\textcolor{OliveGreen}{\:\:+3.7\%} \hspace{0.5cm}  & \hspace{0.5cm} \:\:8.49 \scriptsize\textcolor{OliveGreen}{\:\:+3.9\%} \hspace{0.5cm}\\\hline

L3C \cite{L3C}                    & \hspace{0.5cm} \:\:7.92 \scriptsize\textcolor{OliveGreen}{+13.6\%} \hspace{0.5cm}  & \hspace{0.5cm} \:\:8.82 \scriptsize\textcolor{OliveGreen}{+11.7\%} \hspace{0.5cm}  & \hspace{0.5cm} \:\:9.27 \scriptsize\textcolor{OliveGreen}{+13.5\%} \hspace{0.5cm}\\
RC \cite{Lossy2Lossless}          & \hspace{0.5cm} \:\:7.62 \scriptsize\textcolor{OliveGreen}{\:\:+:9.3\%} \hspace{0.5cm}  & \hspace{0.5cm} \:\:8.79 \scriptsize\textcolor{OliveGreen}{+11.3\%} \hspace{0.5cm}  & \hspace{0.5cm} \:\:9.24 \scriptsize\textcolor{OliveGreen}{+13.1\%} \hspace{0.5cm}\\
Near-Lossless \cite{nearlossless} & \hspace{0.5cm} \:\:7.53 \scriptsize\textcolor{OliveGreen}{\:\:+8.0\%} \hspace{0.5cm}  & \hspace{0.5cm} \:\:7.98$^{*}$ \scriptsize\textcolor{OliveGreen}{+1.0\%} \hspace{0.5cm}  & \hspace{0.5cm} \:\:8.43$^{*}$ \scriptsize\textcolor{OliveGreen}{+3.2\%} \hspace{0.5cm}\\
Ours    & \hspace{0.5cm} \:\:\textbf{6.97} \hspace{0.5cm}   & \hspace{0.5cm} \:\:\textbf{7.90} \hspace{0.5cm} & \hspace{0.5cm} \:\:\textbf{8.17} \hspace{0.5cm} \\ \hline 
\end{tabular}%
}
\label{table:result}
\end{table*}

\subsection{Loss Function}
\label{sec:loss}
LC-FDNet is trained with the following three losses: 1) Error variance loss defined as Eq.~\ref{eq:loss_errorvariance}, 2) reconstruction loss, and 3) bitrate loss.

\noindent \textbf{Reconstruction Loss} We define reconstruction loss as the L1 loss between the ground truth and the predicted subimage:

\begin{equation}\label{eq:loss_reconstruction}
L_{rec} = m_L \cdot \norm{y-\hat{y}_{L}}_1 + m_H \cdot \norm{y-\hat{y}_{H}}_1.
\end{equation}

\noindent Note that we multiply the corresponding frequency mask to the prediction error of LFC and HFC. This lets only the low-frequency components contribute to the reconstruction loss of LFC, and similarly for HFC. This makes the LFC/HFC to be specified for low/high-frequency regions, respectively. Although the reconstruction loss is often neglected in other researches, we figure that adopting this loss leads to stable training and performance enhancement.

\vspace{0.2cm}
\noindent \textbf{Bitrate Loss} Bitrate loss is used to minimize the cross-entropy between the real probability distribution of the quantized residual ($p_{q_L}, p_{q_H}$) and the estimated ($p_L, p_H$), respectively. Formally, it is defined as

\begin{equation}\label{eq:loss_bitrate}
L_{br} = m_L \cdot \norm{-\log p_{L}(q_L)}_1 + m_H \cdot \norm{-\log p_{H}(q_H)}_1.
\end{equation}

\noindent The probability distributions $p_L$ and $p_H$ are trained to classify the corresponding quantized residuals (symbols) $q_L$ and $q_H$ by the cross-entropy loss. This is equivalent to the expected bits per symbol and thus we can directly minimize the coding cost. To restrict the contribution of each frequency component as in the reconstruction loss, we multiply the frequency masks to the corresponding probability distribution.

Altogether, we train our network with the loss:

\begin{equation}\label{eq:loss_total}
L = L_{rec} + \lambda_{br}L_{br} + \lambda_{ev}L_{ev}
\end{equation}

\noindent where $\lambda_{ev}$ and $\lambda_{br}$ are the balancing hyperparmeters. In our experiments, we set both $\lambda_{ev}$ and $\lambda_{br}$ as 1. 

\section{Experiments}

\subsection{Experimental Setup}

\noindent \textbf{Implementation Detail} The detail of the network architecture is provided in the Supplementary Material. For the quantization, we use the round function \textit{i.e.}, $q=round(r)$. The derivative is zero except at integers, which cannot be used in gradient-based optimization. Therefore, we approximate the round function as simple STE~\cite{STE} \textit{i.e.}, $q=r$ in the backward pass since \cite{EnergyCompaction} has shown that different quantization approximation methods have a minor effect on the compression performance. The same problem is introduced when deriving $m_L$ with Eq.~\ref{eq:mask}. This is approximated as $m_L = sigmoid(-(\sigma_y - \tau_y))$ in the backward pass. For our entropy coder, we use ``torchac,'' which is a fast arithmetic coding library for PyTorch developed by the authors of L3C~\cite{L3C}.

\vspace{0.2cm}
\noindent \textbf{Dataset} We validate our method on three benchmark datasets, DIV2K, CLIC.p, and CLIK.m. DIV2K~\cite{DIV2K} is a super-resolution dataset that consists of 2K resolution high-quality images, where 800 images are provided for training and 100 images for evaluation. CLIC mobile (CLIC.m) and CLIC professional (CLIC.p) are datasets released as part of the ``Workshop and Challenge on Learned Image Compression''~\cite{CLIC}. CLIC.m consists of 61 evaluation images which are taken using mobile phones. CLIC.p contains 41 evaluation images which are taken by DSLRs. Most of the images in the CLIC datasets are 2K resolution, but some of them are low resolution as far as $512\times384$.

\vspace{0.2cm}
\noindent \textbf{Training} We train our network with DIV2K training images that are of 2K resolution. We randomly extract a patch of size $128 \times 128$ from the input image during training. Adam optimizer~\cite{Adam} is used for the training, with a batch size of 24 for 3,000 epochs. The learning rate is initially set as $1\times10^{-3}$ and decays by a factor of 0.1 every 1,000 epochs. The training takes 36 hours when trained on a GeForce GTX 1080 Ti.

\vspace{0.2cm}
\noindent \textbf{Evaluation} We compare our method for both learned and non-learned compression algorithms. We compare against the following conventional lossless image codecs: PNG~\cite{PNG}, JPEG-LS~\cite{JPEG-LS}, JPEG2000~\cite{JPEG2000}, WebP~\cite{WebP}, BPG~\cite{BPG}, FLIF~\cite{FLIF} and JPEG-XL~\cite{JPEG-XL}. As for the learned methods, we consider L3C~\cite{L3C}, RC~\cite{Lossy2Lossless}, and Near-Lossless~\cite{nearlossless}. L3C and RC are trained with Open Images dataset~\cite{OpenImages} consisting of 300,000 images. Near-Lossless is trained with the same dataset as our method, the DIV2K dataset. We use \textit{bits per pixel} (bpp) as the evaluation metric, where lower bpp indicates better compression performance.

\subsection{Compression Result}
Table~\ref{table:result} presents the comparisons on the described evaluation sets. It can be seen that our method shows superior performance to both engineered and learning-based codecs. 
Considering DIV2K, our method achieves a 3.2\% gain compared to Near-Lossless, which is also trained with DIV2K. In the case of CLIC.m, non-learning codecs such as FLIF and JPEG-XL outperform existing learning-based methods. Thus, it can be interpreted that learning-based methods are difficult to be generalized to CLIC.m. Nevertheless, our method achieves state-of-the-art performance and outperforms JPEG-XL by 3.3\%. Finally, for CLIC.p, our method shows the best performance achieving 1.0\% gain against Near-Lossless.

In Table~\ref{table:subimage_compression}, we report the compression result for each subimage, \textit{i.e.}, each channel and spatial location. Considering the spatial location, we observe that the compression efficiency enhances in the order of $a \xrightarrow{} d \xrightarrow{} b \xrightarrow{} c$. This is straightforward since more information is supplied as we proceed in the above order. From the perspective of channels, better compression is presented in the order of $Y \xrightarrow{} UV$. This is due to the color transform that reduces the variance in $UV$ channels. Moreover, $V$ channel shows a slight improvement compared to $U$ since we use additional input $U$ when encoding the $V$.

\begin{table}
\centering
\caption{Compression result of each subimage for the DIV2K dataset. Compression performance of subimage $x_{YUV,a}$ is the result of JPEG-XL.}
\resizebox{0.6\columnwidth}{!}{%
\begin{tabular}{c|cccc}
\hline
bpp  & \hspace{0.15cm} $a$ \hspace{0.15cm}  & \hspace{0.15cm} $d$ \hspace{0.15cm} & \hspace{0.15cm} $b$ \hspace{0.15cm} & \hspace{0.15cm} $c$ \hspace{0.15cm}               \\ \hline
$Y$ & -& 0.96 & 0.81 & 0.78 \\ [0.2em]
$U$ & -& 0.59 & 0.45 & 0.45 \\ [0.2em] 
$V$ & -& 0.58 & 0.44 & 0.43 \\ \hline \hhline{-----} 
Total & 2.68 & 2.13 & 1.70 & 1.66 \\ \hline
\end{tabular}%
}
\label{table:subimage_compression}
\end{table}


\begin{figure*}[t]
    \captionsetup[subfigure]{font=large}
    \captionsetup[subfigure]{labelformat=empty}
    \centering

	\begin{subfigure}{0.23\textwidth}
		\includegraphics[width=\linewidth]{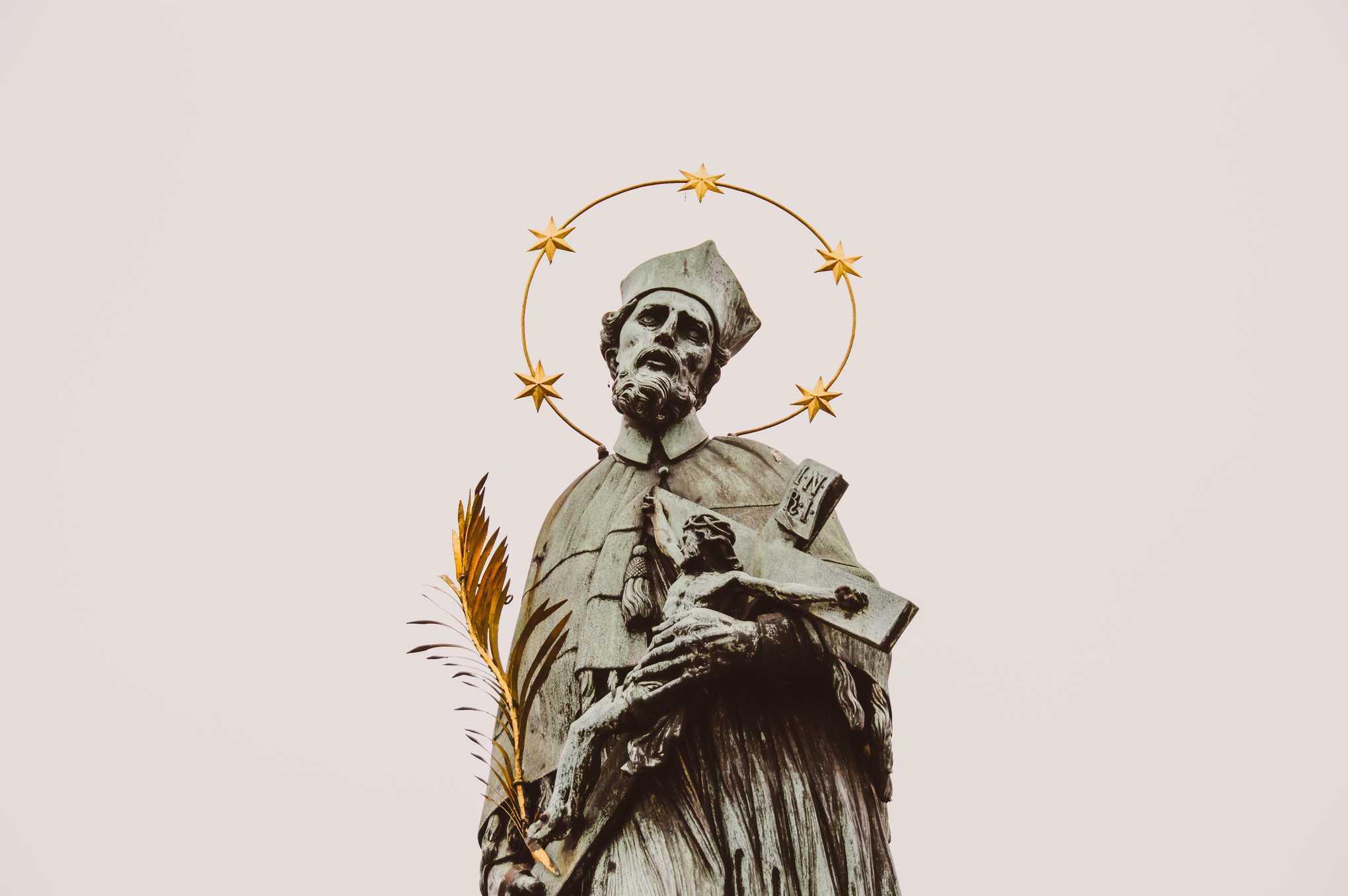}
	\end{subfigure}
	\begin{subfigure}{0.23\textwidth}
		\includegraphics[width=\linewidth]{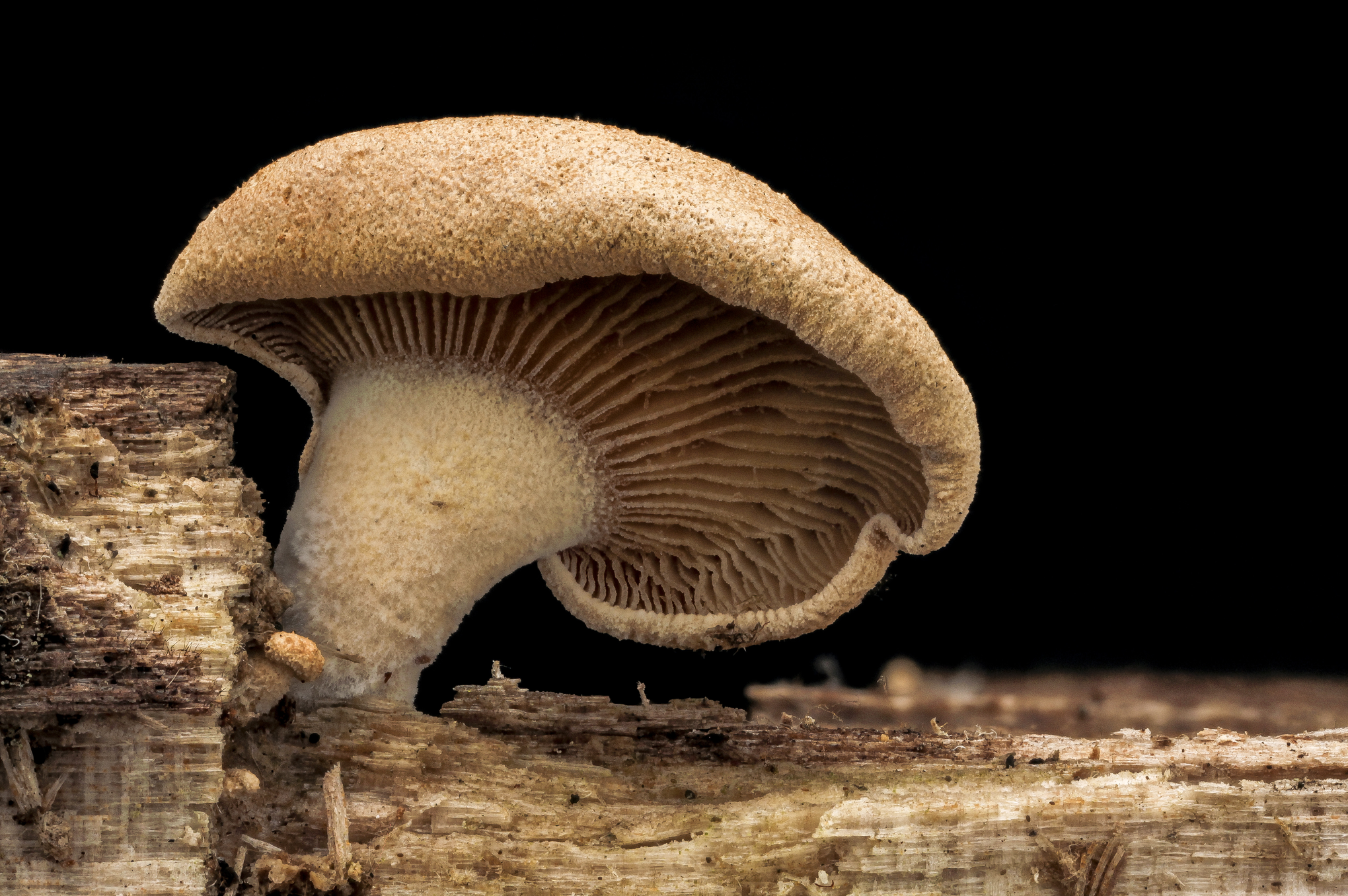}
	\end{subfigure}
	\begin{subfigure}{0.27\textwidth}
		\includegraphics[width=\linewidth]{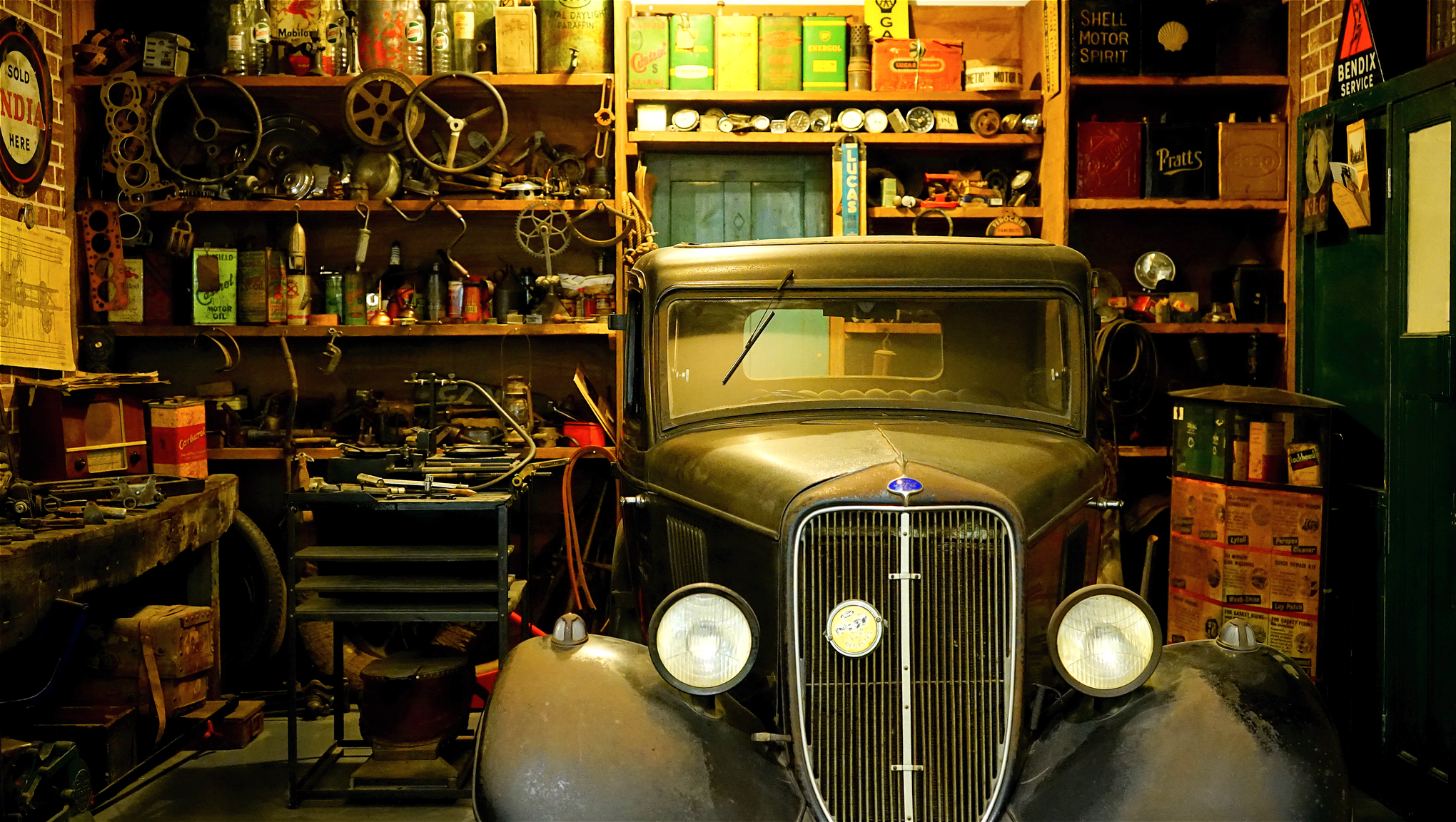}
	\end{subfigure}
	\begin{subfigure}{0.23\textwidth}
		\includegraphics[width=\linewidth]{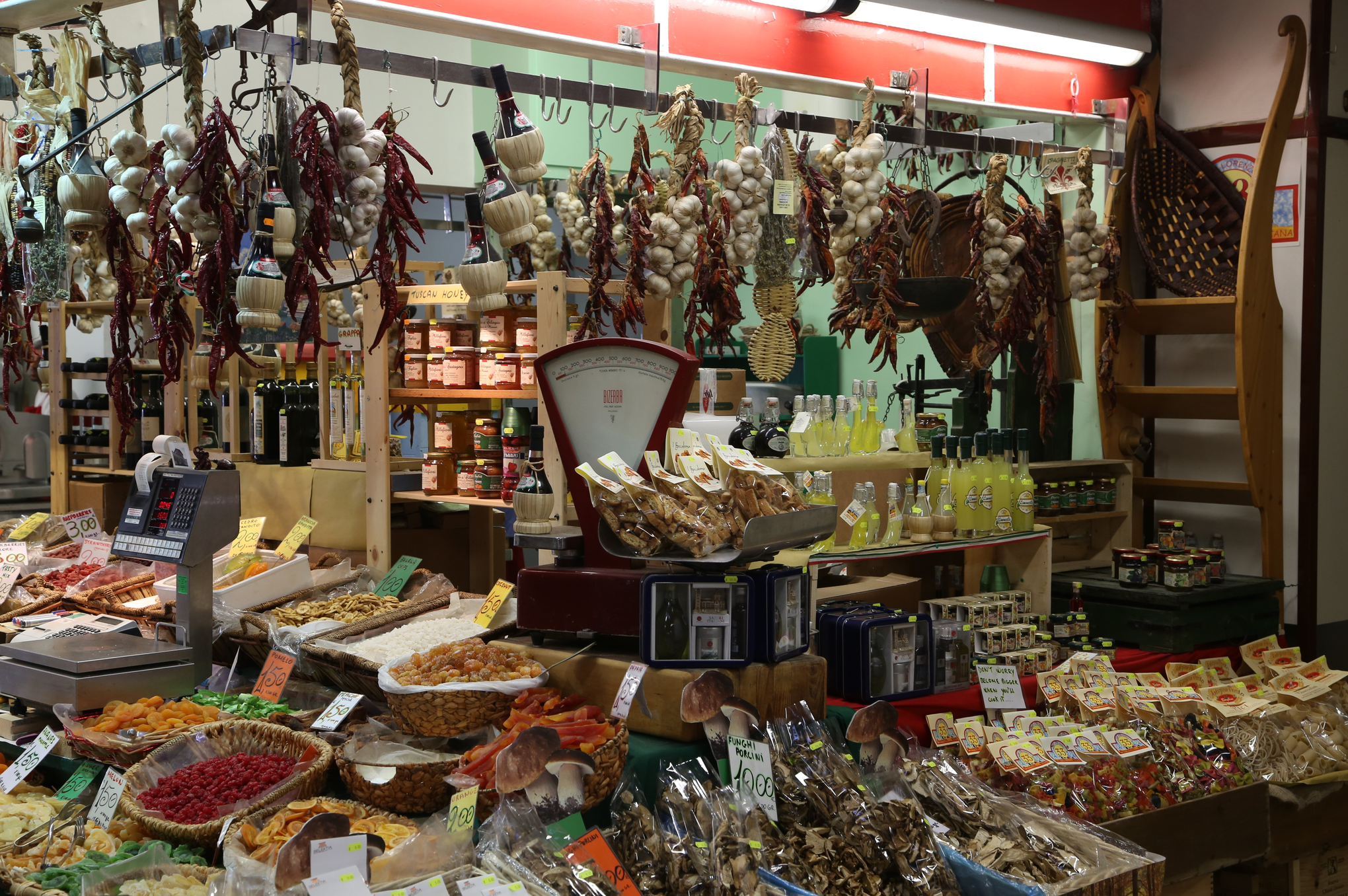}
	\end{subfigure}
    
	\begin{subfigure}{0.23\textwidth}
		\includegraphics[width=\linewidth]{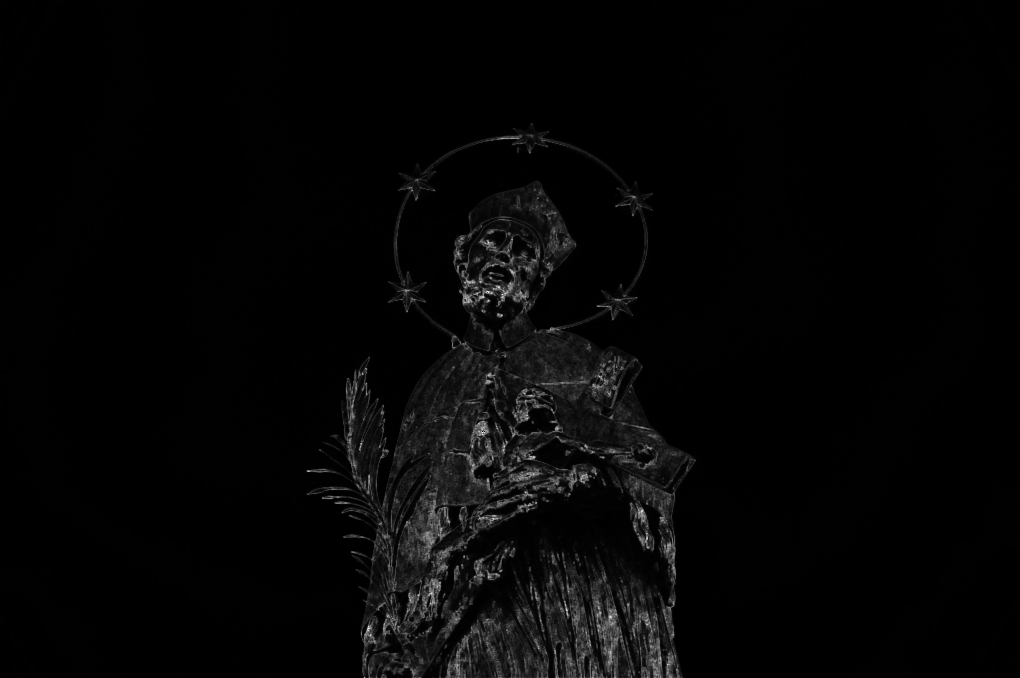}
	\end{subfigure}
	\begin{subfigure}{0.23\textwidth}
		\includegraphics[width=\linewidth]{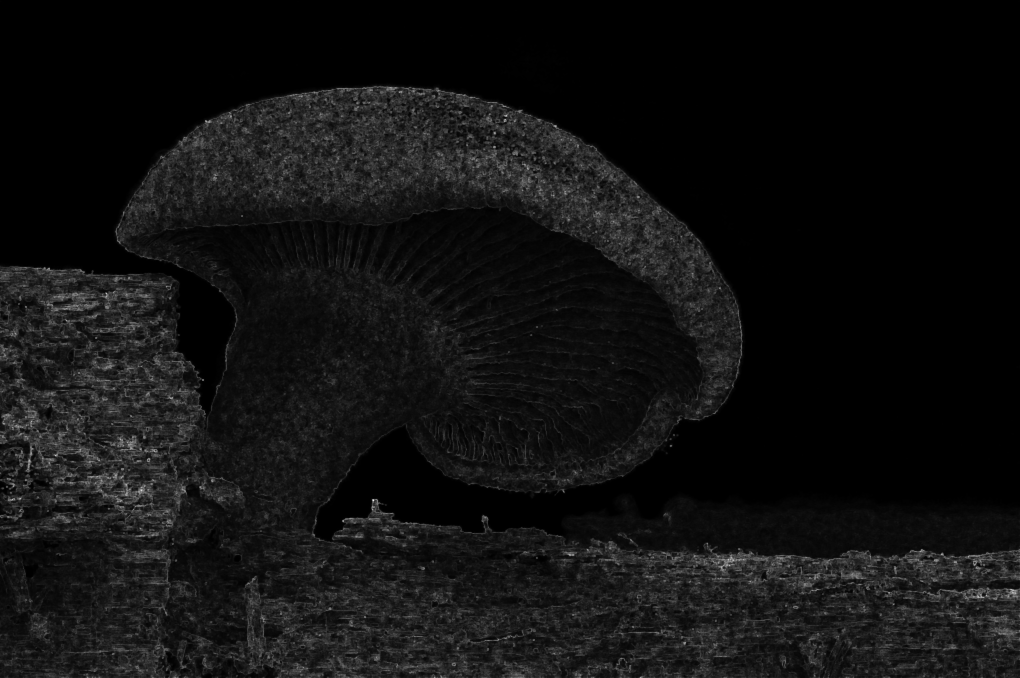}
	\end{subfigure}
	\begin{subfigure}{0.27\textwidth}
		\includegraphics[width=\linewidth]{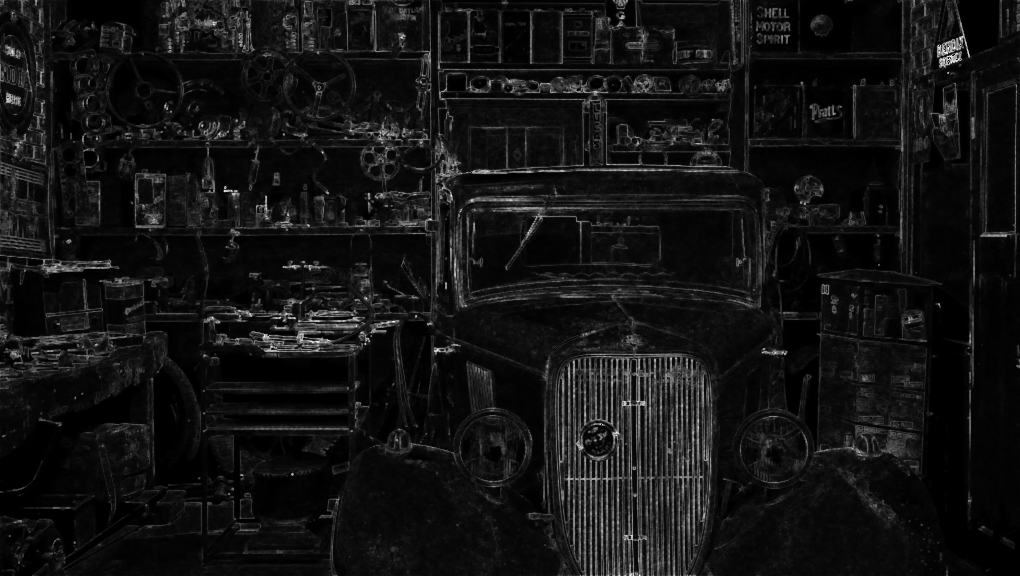}
	\end{subfigure}
	\begin{subfigure}{0.23\textwidth}
		\includegraphics[width=\linewidth]{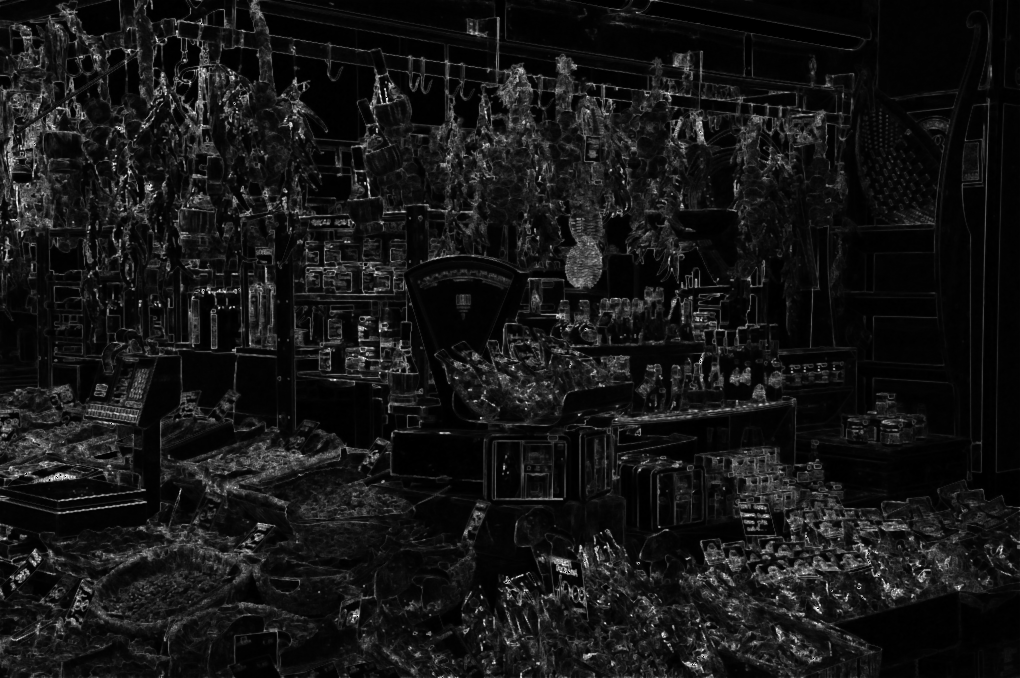}
	\end{subfigure}	

	\begin{subfigure}{0.23\textwidth}
		\includegraphics[width=\linewidth]{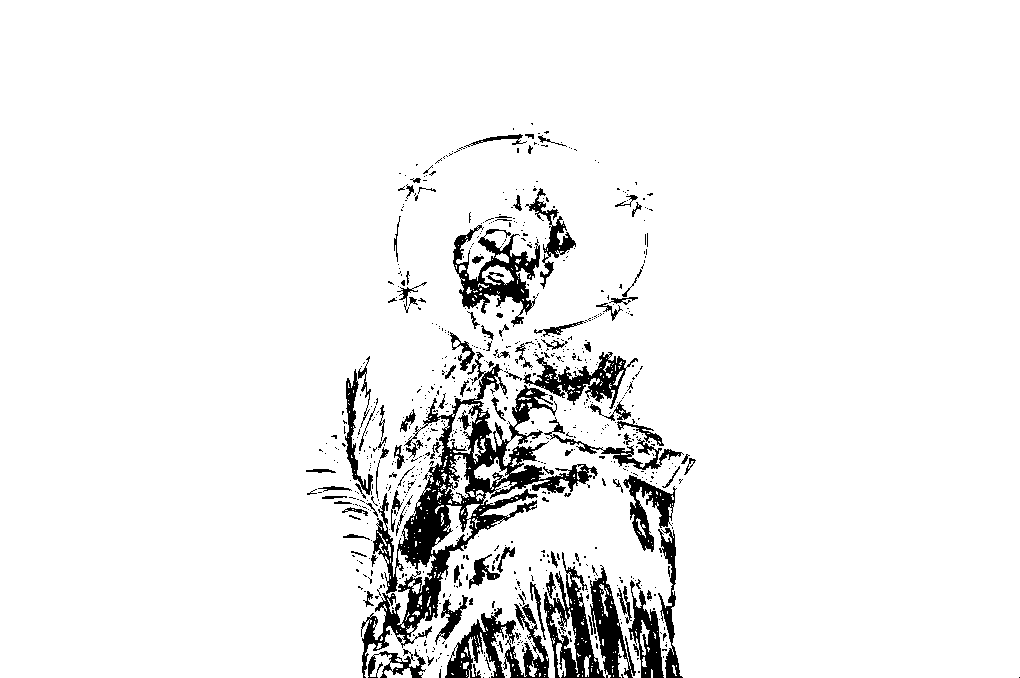}
		\caption{0.26}
	\end{subfigure}
	\begin{subfigure}{0.23\textwidth}
		\includegraphics[width=\linewidth]{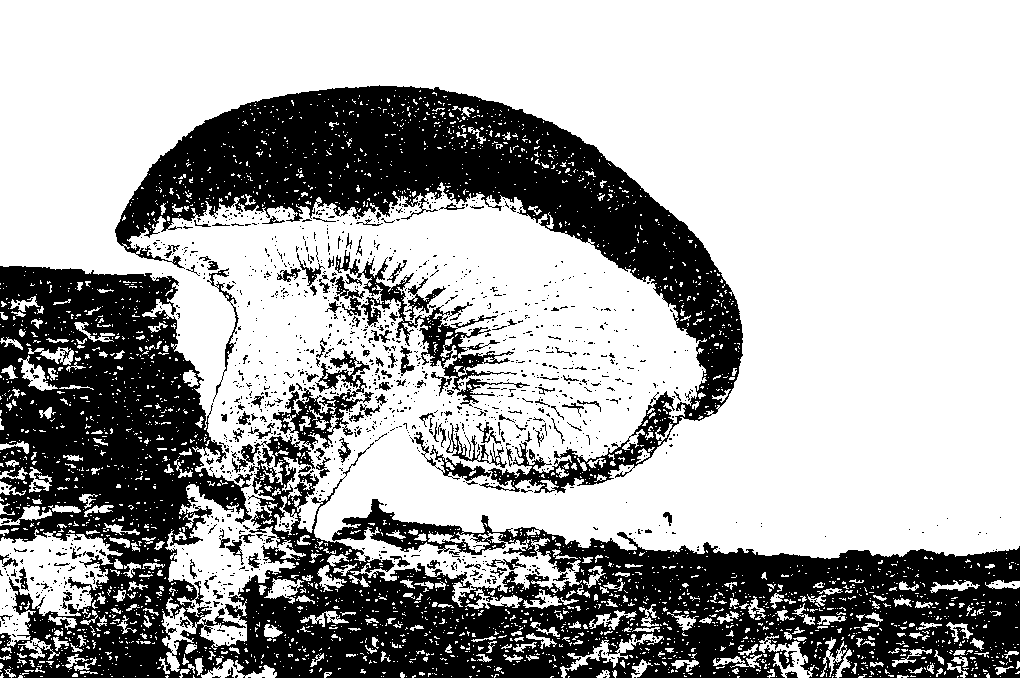}
		\caption{0.34}
	\end{subfigure}
	\begin{subfigure}{0.27\textwidth}
		\includegraphics[width=\linewidth]{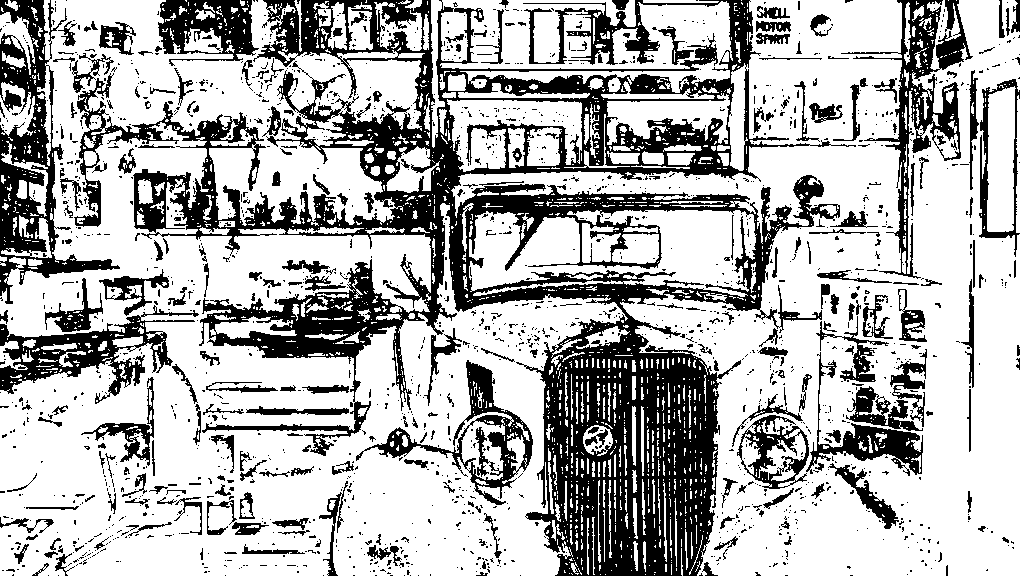}
		\caption{9.05}
	\end{subfigure}
	\begin{subfigure}{0.23\textwidth}
		\includegraphics[width=\linewidth]{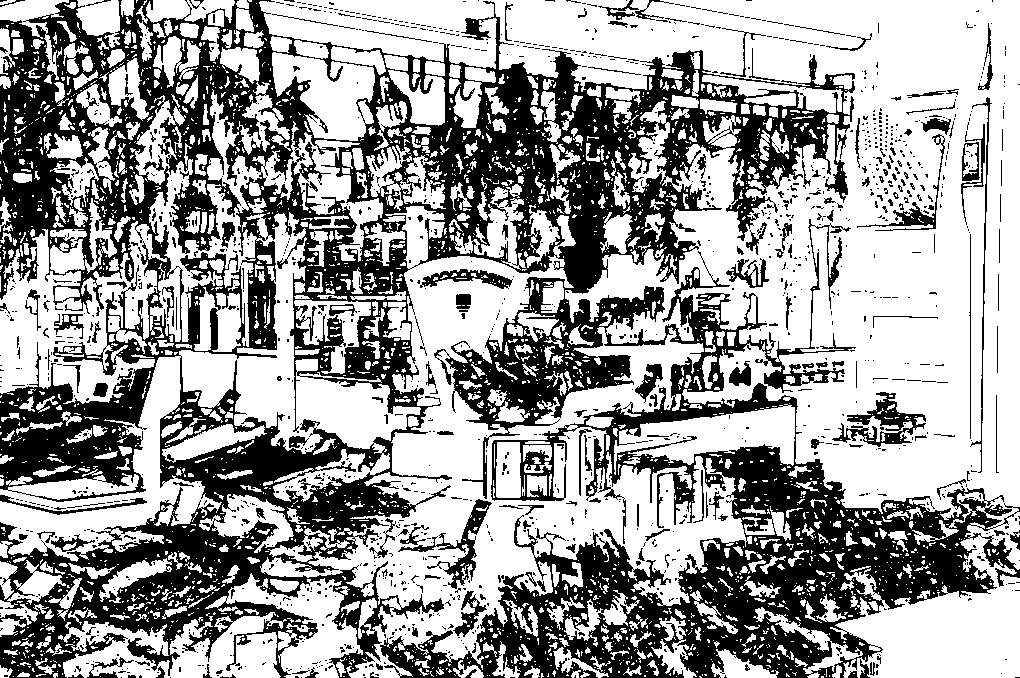}
		\caption{9.33}
	\end{subfigure}	

	\caption{From top to down are the visualizations of an input image, error variance map, low-frequency mask, and error variance threshold. These elements are visualized for the case of the $Y$ channel and $d$ location. We choose the samples from the DIV2K dataset that have the smallest and largest $\tau_y$. The error variance map is magnified by 5 for visualization.}  
	\label{fig:element_visualization}	
\end{figure*}

\subsection{Inference Time} 
We measure the inference time required for encoding a 512$\times$512 image on an GeForce GTX 1080 Ti. First, the compression of the initial subimage using JPEG-XL requires 199\:ms. The forward pass for achieving the quantized residual, probability distribution, and frequency mask takes 33\:ms. Finally, the arithmetic coding using torchac requires 1.6\:s. In total, our method requires 1.8\:s. Note that 89\% of the time is consumed in arithmetic coding, which can be shortened if PyTorch-friendly entropy coder is developed in the future.

\subsection{AFD Analysis}
We quantitatively and qualitatively demonstrate that the error variance threshold is adaptive to the channel, spatial location, and image characteristics. We first show that the error variance threshold is adaptive to the channel and spatial location through Table~\ref{table:subimage_threshold}. In general, the threshold value decreases in the order of $d \xrightarrow{} b \xrightarrow{} c$ and $Y \xrightarrow{} UV$. This is consistent with the order of compression efficiency. If the subimage is more predictable, the overall values in the variance map tend to decrease. In this case, the error variance threshold should also decrease to balance the low to high-frequency ratio.

Next, we show that the error variance threshold is adaptive to the image characteristics. Fig.~\ref{fig:element_visualization} shows the outputs generated from LFC. The first two samples contain a large portion of the smooth background and a single object. These samples produce a small threshold value of 0.26 and 0.34. In contrast, the last two samples are more complicated than the preceding ones and introduce many high-frequency components. These samples generate a large threshold value of 9.05 and 9.33. We interpret these observations that the error variance threshold is proportional to the number of high-frequency components an image contains. 

\begin{figure}
    \centering
    \begin{tabular}{cc}
        \includegraphics[width=0.48\columnwidth]{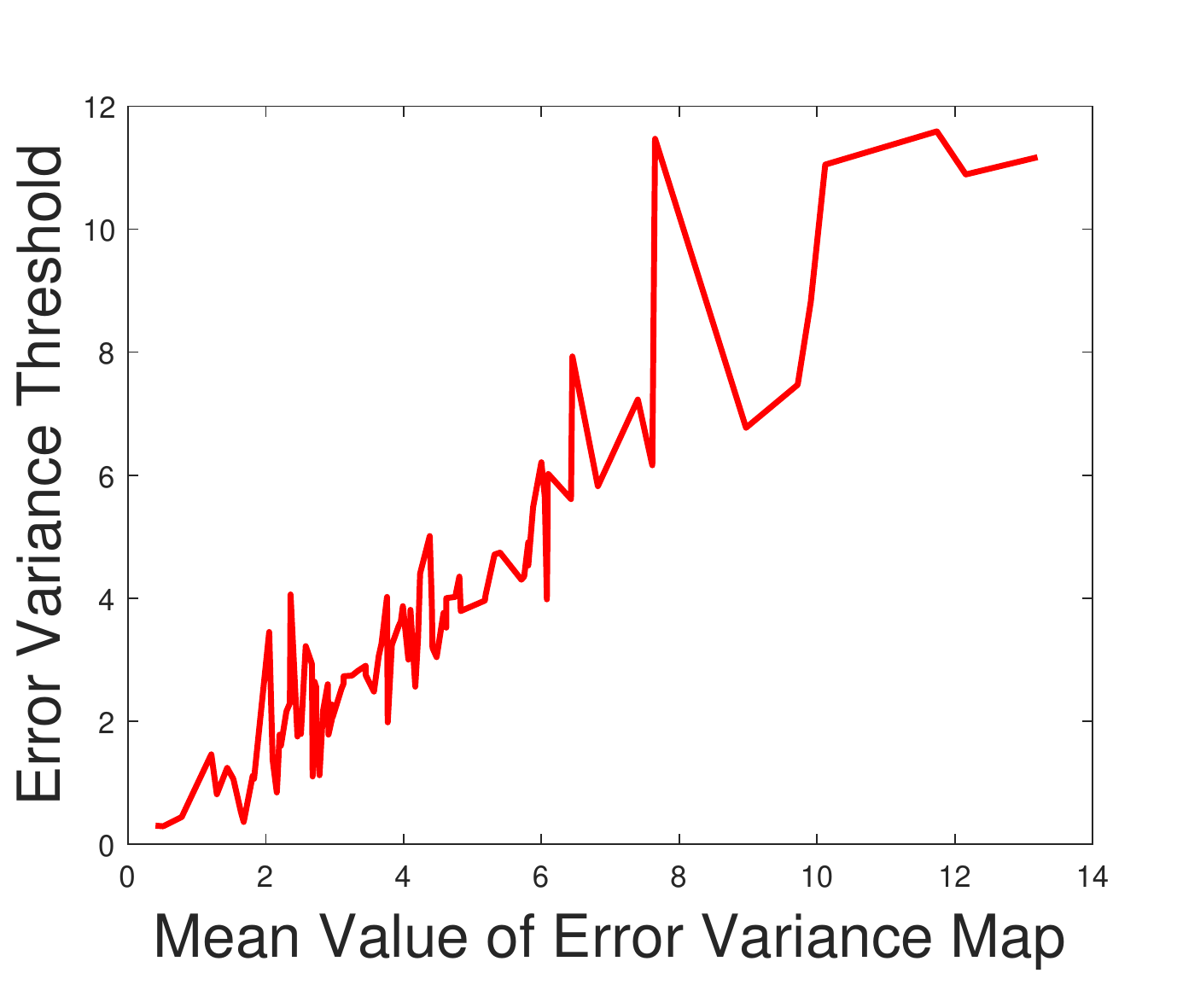}&
        \includegraphics[width=0.48\columnwidth]{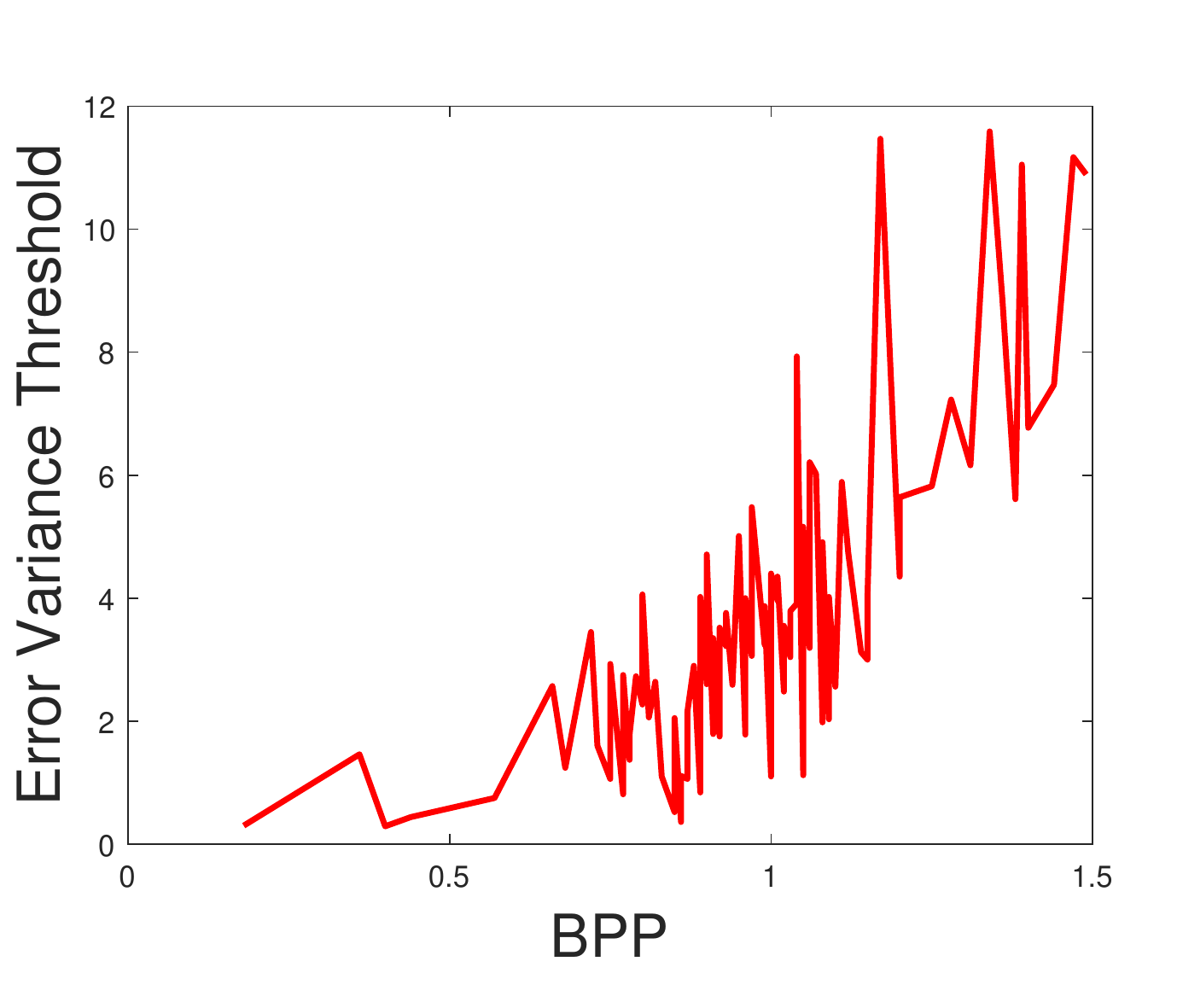}\\ 
        
    \end{tabular}
	\caption{Graph of image characteristic versus error variance threshold for DIV2K dataset. We use the samples of DIV2K, channel of $Y$, and spatial location of $d$.}
	\label{fig:correlation}    
\end{figure}

We also verify the above conclusion quantitatively. We figure that images with many high-frequency components tend to introduce large values in the error variance map. In addition, these images result in low compression rate. Hence, for samples in DIV2K, we plot the error variance threshold against the mean value of the error variance map and bpp in Fig.~\ref{fig:correlation}. It can be observed that the error variance threshold and the two components have a positive correlation. In conclusion, the error variance threshold is adaptive to image characteristics, where the threshold value increases as more high-frequency components are present in the image.

\vspace{0.8cm}

\begin{table}
\centering
\caption{Error variance threshold value for each subimage. Since the threshold is image-specific, we average the threshold for all the images in DIV2K dataset.}
\resizebox{0.55\columnwidth}{!}{%
\begin{tabular}{c|ccc}
\hline
\hspace{0.25cm} $\tau_y$ \hspace{0.25cm}    & \hspace{0.25cm} $d$ \hspace{0.25cm}   & \hspace{0.25cm} $b$ \hspace{0.25cm}           & \hspace{0.25cm} $c$ \hspace{0.25cm}         \\ [0.1em] \hline
 $Y$ & 3.57 & 2.84 & 2.72\\ [0.2em]
$U$ & 2.68 & 2.24 & 2.21 \\ [0.2em]
$V$ & 2.66 & 2.30 & 2.24 \\ \hline
\end{tabular}%
}
\label{table:subimage_threshold}
\end{table}

\subsection{Ablation Study}
Several ablation experiments are performed to analyze each component of LC-FDNet. We demonstrate the contribution of each component in Table~\ref{table:ablation} by excluding the components one by one. The comparing networks are trained and evaluated on the DIV2K dataset. We exclude the portion of JPEG-XL\:(2.68\:bpp) in computing the compression performance.

\vspace{0.2cm}
\noindent \textbf{Coarse to Fine} We first demonstrate the effect of proceeding in a coarse-to-fine manner. We design a comparing network that compresses the low and high-frequency components together so that the high-frequency components do not benefit from the low ones. Hence, the network only outputs the subimage prediction and probability distribution. We match the number of parameters for both networks to demonstrate that the performance gain does not come from the difference in the network size. The result (first row of Table~\ref{table:ablation}) shows that we can have a 5.8\% performance gain by the coarse-to-fine processing.  Hence, we can conclude that low-frequency components act as strong priors for estimating the high-frequency components.

\vspace{0.2cm}
\noindent \textbf{Adaptive Error Variance Threshold} We show that letting the error variance threshold be adaptive to image characteristics leads to performance enhancement. Specifically, we train a network with fixed $\tau$ for every subimage. Since our framework is sensitive to the value of $\tau$, we should carefully set the threshold value for a fair comparison. Hence, we use the average $\tau_y$ of the DIV2K validation set derived from our full model as our fixed $\tau$. The second row of the table shows that the compression efficiency decreases by 2.7\% with the fixed $\tau$.

\vspace{0.2cm}
\noindent \textbf{Loss Masking} We verify that multiplying the corresponding frequency mask in Eq.~\ref{eq:loss_reconstruction} and Eq.~\ref{eq:loss_bitrate} has a valid contribution. For this, we train a network without the multiplication of frequency mask. In other words, the LFC and HFC of this network share the same objective and are not frequency-specific. In this scenario, a performance drop of 2.0\% is observed as in the third row, indicating that assigning frequency-specific roles to LFC and HFC has a positive influence.

\subsection{Frequency Component Analysis}
We show the performance enhancement in low and high-frequency regions separately in Table~\ref{table:C2F} for further analysis of our system. We first compare our full model (C2F) against the model without the coarse-to-fine processing (w/o C2F) as in the ablation study. Compared to the network that proceeds without the coarse-to-fine processing, the low-frequency components have a 3.4\% performance gain, whereas high-frequency components have an 11.3\% increase. This implies that the high-frequency components benefit significantly from the coarse-to-fine processing. The low-frequency components indeed act as strong conditions for the estimation of high-frequency components, as intended.

We train an additional network that proceeds in a fine-to-coarse manner (F2C). That is, we compress the high-frequency components first and utilize them for encoding the low-frequency components. From Table~\ref{table:C2F}, we observe that the performance gain is 0.5\% in the low-frequency area, which is minor. In contrast, there is a considerable performance drop of 10.8\% in the high-frequency components. Altogether, there is a total performance drop of 2.9\% when proceeding in a fine-to-coarse manner. Although low-frequency components take up a large portion of an image, the gain is too small to have enough contribution to the overall gain. Thus, we conclude that the design choice of coarse-to-fine manner is indeed favorable.

\begin{table}
\centering
\caption{Ablation study of our method on DIV2K dataset. C2F refers to the coarse-to-fine network. \checkmark \, indicates that the corresponding element is used.}
\resizebox{0.95\columnwidth}{!}{%
\begin{tabular}{ccc|l}
\hline
\hspace{0.2cm} C2F \hspace{0.2cm}          & Adaptive $\tau$         & Loss Masking  & bpp \\ \hline
             & \checkmark &            & 5.81 \scriptsize\textcolor{OliveGreen}{+5.8\%} \\
\checkmark   &            & \checkmark & 5.64 \scriptsize\textcolor{OliveGreen}{+2.7\%} \\ 
\checkmark   & \checkmark &            & 5.60 \scriptsize\textcolor{OliveGreen}{+2.0\%} \\
\checkmark   & \checkmark & \checkmark & 5.49 \\  \hline
\end{tabular}%
}
\label{table:ablation}
\end{table}

\begin{table}
\centering
\caption{Performance gain on low and high-frequency regions. F2C refers to the network proceeding in a fine-to-coarse manner.}
\resizebox{0.90\columnwidth}{!}{%
\begin{tabular}{c|ccc}
\hline
Method   & Low-Freq & High-Freq & Total \:\:\:\:\:\:\:\:\: \\ \hline
w/o C2F     & 3.95 \scriptsize\textcolor{OliveGreen}{+3.4\%} & 1.86 \scriptsize\textcolor{OliveGreen}{+11.3\%} & 5.81 \scriptsize\textcolor{OliveGreen}{+5.8\%}\\
F2C & 3.80 \scriptsize\textcolor{Red}{-0.5\%} & 1.85 \scriptsize\textcolor{OliveGreen}{+10.8\%} & 5.65 \scriptsize\textcolor{OliveGreen}{+2.9\%} \\
C2F & 3.82 \:\:\:\:\:\:\:\: & 1.67 \:\:\:\:\:\:\:\:\: & 5.49 \:\:\:\:\:\:\:\: \\ \hline
\end{tabular}%
}
\label{table:C2F}
\end{table}

\section{Conclusion}
We have proposed LC-FDNet, a lossless image compression framework that decomposes an input image into low and high-frequency regions to proceed in a coarse-to-fine manner. We resolved the performance drop in the high-frequency areas by first compressing the low-frequency components and using them as a strong prior for encoding the remaining high-frequency components. Furthermore, we designed the frequency decomposition method to be adaptive to color channel, spatial location, and image characteristics to derive the image-specific optimal ratio of low/high-frequency components. Extensive experiments show that our method achieves state-of-the-art performance for high-resolution benchmark datasets. We will release our code publicly.

{\small
\bibliographystyle{ieee_fullname}
\bibliography{egbib}
}

\end{document}